\documentclass[aps,pre,reprint,showpacs,superscriptaddress,longbibliography]{revtex4-1}
\usepackage{mathtools,amssymb,graphicx,units}
\usepackage[usenames,dvipsnames]{color}

\usepackage[plainpages=false,pdfpagelabels,colorlinks=true,linkcolor=red,urlcolor=blue,citecolor=blue,pdftitle={Title},pdfauthor={},pdfdisplaydoctitle=true,pdfduplex=DuplexFlipLongEdge]{hyperref}

\renewcommand{\vec}[1]{\boldsymbol{#1}}

\begin{document}
\title{Eigenstate thermalization in the two-dimensional transverse field Ising model:\\ II. Off-diagonal matrix elements of observables}
\author{Rubem Mondaini}
\affiliation{Beijing Computational Science Research Center, Beijing 100193, China} 
\author{Marcos Rigol}
\affiliation{Department of Physics, Pennsylvania State University, University Park, Pennsylvania 16802, USA}

\begin{abstract}
We study the matrix elements of few-body observables, focusing on the off-diagonal ones, in the eigenstates of the two-dimensional transverse field Ising model. By resolving all  symmetries, we relate the onset of quantum chaos to the structure of the matrix elements. In particular, we show that a general result of the theory of random matrices, namely, the value 2 of the ratio of variances (diagonal to off-diagonal) of the matrix elements of Hermitian operators, occurs in the quantum chaotic regime. Furthermore, we explore the behavior of the off-diagonal matrix elements of observables as a function of the eigenstate energy differences, and show that it is in accordance with the eigenstate thermalization hypothesis ansatz.
\end{abstract}

\maketitle

\section{Introduction}

Whether quantum statistical behavior can emerge in isolated systems, which evolve unitarily as dictated by quantum mechanics, is a question that has fascinated physicists since the early days of quantum mechanics \cite{vonneumann_10}. It has taken almost a century for experimental setups with the degree of isolation and control required to address such a question to become widely available \cite{bloch_dalibard_review_08, cazalilla_citro_review_11}. Those experimental setups, involving mostly ultracold gases trapped in ultrahigh vacuum, have begun to be used to study quantum thermalization (or the lack thereof) in a variety of settings \cite{kinoshita_wenger_06, gring_kuhnert_12, trotzky_chen_12, langen_erne_15, schreiber_hodgman_15, kaufman_tai_16, clos_porras_16, choi_hild_16, smith_lee_16, neill_roushan_16}.

In parallel with the experimental activity, computational studies have shown that equilibration of observables can occur in quantum systems even if their dynamics are unitary \cite{rigol_dunjko_07, rigol_muramatsu_06, kollath_lauchli_07, manmana_wessel_07, rigol_dunjko_08, cramer_flesch_08a, cramer_flesch_08b, rigol_09a, rigol_09b, eckstein_kollar_09, rigol_santos_10, banuls_cirac_11, calabrese_essler_11, gramsch_rigol_12, khatami_pupillo_13, zangara_dente_13, wright_rigol_14, sorg_vidmar_14, fagotti_collura_14, bertini_essler_15, brandino_caux_15, balz_reimann_17}, but only those that are nonintegrable (quantum chaotic) are generally described by traditional statistical mechanics after equilibration \cite{rigol_dunjko_07, rigol_16}. This can be understood in the context of the eigenstate thermalization hypothesis (ETH) \cite{deutsch_91, srednicki_94, rigol_dunjko_07} (see Ref.~\cite{dalessio_kafri_16} for a recent review), which states that the diagonal matrix elements of observables in the eigenstates of the Hamiltonian are smooth functions of the energy~\footnote{The differences between the diagonal matrix elements of observables in neighbor eigenstates are exponentially small in the system size}, while the off-diagonal ones are exponentially small in the system size. For an observable $\hat O$, the ETH ansatz can be written as~\cite{srednicki_99, dalessio_kafri_16}
\begin{equation}
 \label{eq:eth_ansatz}
 O_{\alpha\beta} = {\cal O}(\bar E)\delta_{\alpha\beta} + e^{-S(\bar E)/2}f_{O}(\bar E, \omega)R_{\alpha\beta},
\end{equation}
where $\bar E \equiv (E_\alpha+E_\beta)/2$, $S(E)$ is the thermodynamic entropy at energy $E$, and $\omega\equiv E_\alpha-E_\beta$. $\cal O$ and $f_{O}$ are smooth functions of their arguments, while $R_{\alpha\beta}$ are random numbers with zero mean and unit variance. The connection with statistical mechanics is immediate through ${\cal O}(E)$, which is the statistical mechanics prediction for $\hat O$ at the mean energy $E$. Eigenstate thermalization has been observed in a variety of nonintegrable lattice models \cite{rigol_dunjko_08, rigol_09a, rigol_09b, rigol_santos_10, biroli_kollath_10, roux_2010, neuenhahn_marquardt_12, khatami_pupillo_13, steinigeweg_herbrych_13, sorg_vidmar_14, beugeling_moessner_14, kim_ikeda_14, steinigeweg_khodja_14, khodja_steinigeweg_15, beugeling_moessner_15, fratus_srednicki_15, mondaini_fratus_16, luitz_16, luitz_17}.

To understand how the ETH ansatz~\eqref{eq:eth_ansatz} explains thermalization in isolated quantum systems, let us consider an initial state $|\Psi_I\rangle=\sum_\alpha c_\alpha |\alpha\rangle$, where $|\alpha\rangle$ are the eigenstates of the Hamiltonian $\hat{H}$ generating the dynamics ($\hat{H}|\alpha\rangle=E_\alpha|\alpha\rangle$). Using that $|\Psi(t)\rangle\equiv\exp[-{\rm i}\hat{H}t]|\Psi_I\rangle=\sum_\alpha c_\alpha \exp[-{\rm i}E_\alpha t]|\alpha\rangle$, we set $\hbar=1$, the time evolution of $\hat O$, $O(t) \equiv \langle\Psi(t)|\hat O|\Psi(t)\rangle$, follows from
\begin{equation}
 O(t) = \sum_\alpha |c_\alpha|^2 O_{\alpha\alpha} +\sum_{\substack{\alpha,\beta \\ (\alpha\neq\beta)}} c_\alpha^* c_\beta e^{{\rm i}(E_\alpha-E_\beta)t}O_{\alpha\beta}.
 \label{eq:O_t}
\end{equation}
The fact that off-diagonal matrix elements of observables are exponentially small in the system size ensures that, after dephasing, the second term in Eq.~\eqref{eq:O_t} is exponentially smaller than the first term. The smooth dependence of $O_{\alpha\alpha}$ on the energy, combined with the fact that the width of the energy distribution of physical initial states (such as those in quantum quenches involving local Hamiltonians \cite{rigol_dunjko_08}) is generally subextensive as in statistical mechanics ensembles, then ensures that the expectation values of observables after relaxation are the ones predicted by statistical mechanics \cite{dalessio_kafri_16}.

While the expectation values of observables after equilibration are determined by diagonal matrix elements, the first term in Eq.~\eqref{eq:O_t}, the dynamics that results in equilibration and the fluctuations of observables about their equilibrated values are determined by the exponentially small off-diagonal matrix elements (and the initial state), the second term in Eq.~\eqref{eq:O_t}. Also, the fact that fluctuation-dissipation relations hold in isolated quantum-chaotic systems without the need of assuming thermal equilibrium is encoded in the structure of the off-diagonal matrix elements of observables \cite{khatami_pupillo_13, dalessio_kafri_16}. Early studies of eigenstate thermalization explored the qualitative behavior of off-diagonal matrix elements of observables in near-integrable and quantum-chaotic regimes~\cite{rigol_dunjko_08, rigol_09b, santos_rigol_10b}, while more recent studies have looked into their distribution and scaling of their magnitude with system size, as well as the functional form of $f_{O}(\bar E, \omega)$, in one-dimensional chains \cite{khatami_pupillo_13, steinigeweg_herbrych_13, beugeling_moessner_15, dalessio_kafri_16, luitz_17}. Still, the off-diagonal matrix elements of observables remain much less studied than the diagonal ones.

Here, we study the behavior of the off-diagonal matrix elements of few-body observables, such as the structure factor and nearest neighbor spin correlations, in the eigenstates of the two-dimensional transverse field Ising model (2D-TFIM) on the square lattice. This work extends the analysis of the same model in Ref.~\cite{mondaini_fratus_16}, in which we studied quantum chaos indicators and the diagonal matrix elements of observables. The 2D-TFIM exhibits quantum-chaotic behavior, and the diagonal matrix elements of few-body observables comply with the ETH ansatz~\eqref{eq:eth_ansatz}, for nonvanishing but finite values of the field. We note that, despite its simplicity, the 2D-TFIM hosts both a zero-temperature quantum phase transition with increasing the strength of the transverse field, and a finite-temperature one, separating an ordered and a paramagnetic phase. Remarkably, the one-dimensional TFIM was recently realized experimentally using ultracold bosonic atoms in tilted optical lattices~\cite{simon_bakr_11}.

One of our main goals in this work is to show that, in the quantum-chaotic regime, the matrix elements of observables in the 2D-TFIM satisfy a striking prediction from the Gaussian orthogonal ensemble (GOE) of random matrix theory. Namely, that the ratio between the variance of diagonal and off-diagonal matrix elements of observables in very small energy windows is universal and equal to 2 \cite{dalessio_kafri_16}. We also study the scaling of the magnitude of off-diagonal matrix elements of observables, and explore the existence and behavior of the function $f_{O}(\bar E, \omega)$ introduced in the ETH ansatz \eqref{eq:eth_ansatz}.

The presentation is organized as follows. In Sec.~\ref{sec:sec2}, we introduce the model and the numerical methods used. Section~\ref{sec:sec3} explores the connection between the onset of quantum chaos and the GOE result for the ratio between variances of the diagonal and off-diagonal matrix elements of observables. The scaling and behavior of the off-diagonal matrix elements, and their relation to the ETH ansatz, are studied in Sec.~\ref{sec:sec4}. A summary of our results is presented in Sec.~\ref{sec:sec5}.

\section{Model and Numerical Method}\label{sec:sec2}
The Hamiltonian of the 2D-TFIM, assuming periodic boundary conditions, reads
\begin{equation}
 \hat H = -J\sum_{\langle {\bf i},{\bf j}\rangle}\hat\sigma_{\bf i}^z\hat\sigma_{\bf j}^z + g\sum_{\bf i}\hat\sigma_{\bf i}^x,
 \label{eq:hamiltonian}
\end{equation}
where $\hat\sigma_{\bf i}^z$ ($\hat\sigma_{\bf i}^x$) is the $z$ ($x$) Pauli matrix at site ${\bf i}$ of the lattice. The strength of the nearest neighbor (denoted by $\langle {\bf i},{\bf j}\rangle$ in the constrained summation) Ising exchange interactions is given by $J$. In this work, we focus on the ferromagnetic case $(J>0)$, setting $J=1$ as our energy scale. (In Ref.~\cite{mondaini_fratus_16}, we considered both the ferromagnetic and antiferromagnetic cases.) Lastly, $g$ denotes the magnitude of the transverse field. The majority of the results presented refer to regular square lattices with number of sites $N = \ell_x \times \ell_y$, where $\ell_{x}$ and $\ell_{y}$ denote the linear dimensions of the system in the $x$ and $y$ directions, respectively. We also solve for a tilted lattice with 20 sites, one of the two largest lattices (lattice 20A) studied in Ref.~\cite{mondaini_fratus_16}. Here, the largest lattice considered is a regular lattice with 25 sites, while the smallest one, also regular, has 15 sites. We study various system sizes to carry out finite-size scaling analyses of observables of interest.

The 2D-TFIM [Eq.~(\ref{eq:hamiltonian})] on the 2D square lattice possesses a variety of symmetries, which once identified and taken into account allow one to block diagonalize the Hamiltonian. Subsectors of the Hamiltonian with no symmetries are needed to establish the existence of quantum chaotic behavior using the energy spectrum \cite{santos_rigol_10a, santos_rigol_10b, mondaini_fratus_16}. We make use of the following symmetries: translation ($\hat T$), spin-flipping ($\hat Z_2$), mirror in $x$ ($\hat S_x$), mirror in $y$ ($\hat S_y$) and, when applicable, mirror along the $x=y$ line ($\hat S_{xy}$). The latter symmetry is present when $\ell_x = \ell_y$ and the parity is the same under $\hat S_x$ and $\hat S_y$. After applying these symmetries, the subsectors of the Hamiltonian are diagonalized using full exact diagonalization.

In the best case scenario, when all those symmetries are present, the original Hilbert space (of dimension ${\cal D} = 2^N$) can be split into smaller subsectors, some of which have dimension ${\cal D}^\prime \approx \frac{2^{N-4}}{N}$. As discussed in  Appendix~\ref{sec:appendix_split}, the largest subsector we fully diagonalize has $166,752$ states. It corresponds to the zero-momentum subsector of the $5\times5$ lattice, after applying the $\hat Z_2,\, \hat S_x,$ and $\hat S_y$ operations, and for parities under $\hat S_x$ and $\hat S_y$ obeying $\lambda_{\hat S_x}\cdot\lambda_{\hat S_y} = -1$. For regular lattices we focus on the zero momentum subsector [$\vec k = (0,0)$], while for the 20-sites tilted lattice we focus on the $\vec k = \frac{\pi}{5}(2,1)$ momentum subsector, in which the only remaining symmetry to be resolved is $\hat Z_2$.

\section{Quantum chaos indicators}\label{sec:sec3}

\subsection{Ratio of adjacent gaps}

We begin studying the so-called rigidity of the spectrum, i.e., by checking whether level repulsion takes place~\cite{bohigas_giannoni_84, brody_flores_81, haake_91}. This is a fundamental insight from random matrix theory, which has been shown to apply to a variety of nonintegrable lattice models~\cite{dalessio_kafri_16}. 

Given that the 2D-TFIM is integrable in both the classical $(g\to0)$ and paramagnetic $g\to\infty$ (one-site) limits, we test level repulsion when $g$ and $J$ have the same magnitude, $g = J = 1$. Specifically, we compute the ratio of adjacent energy gaps~\cite{oganesyan_huse_07, atas_bogomolny_13}, $r_n \equiv \min\left(\delta_{n+1},\delta_{n}\right)/\max\left(\delta_{n+1},\delta_{n}\right)$, where $\delta_n = E_{n}-E_{n-1}$ is the difference between consecutive energy levels, and $\{E_n\}$ is the sorted list of eigenenergies (from the lowest to the highest) in each subsector of the Hamiltonian. Here is where the necessity of resolving all symmetries becomes apparent: subsectors of the Hamiltonian in which there are unresolved symmetries exhibit uncorrelated energy levels (and possibly extensive degeneracies), i.e., level repulsion is absent even if the Hamiltonian is quantum chaotic~\cite{santos_rigol_10a, santos_rigol_10b, mondaini_fratus_16}.

\begin{figure}[!tb] 
 \includegraphics[width=0.99\columnwidth]{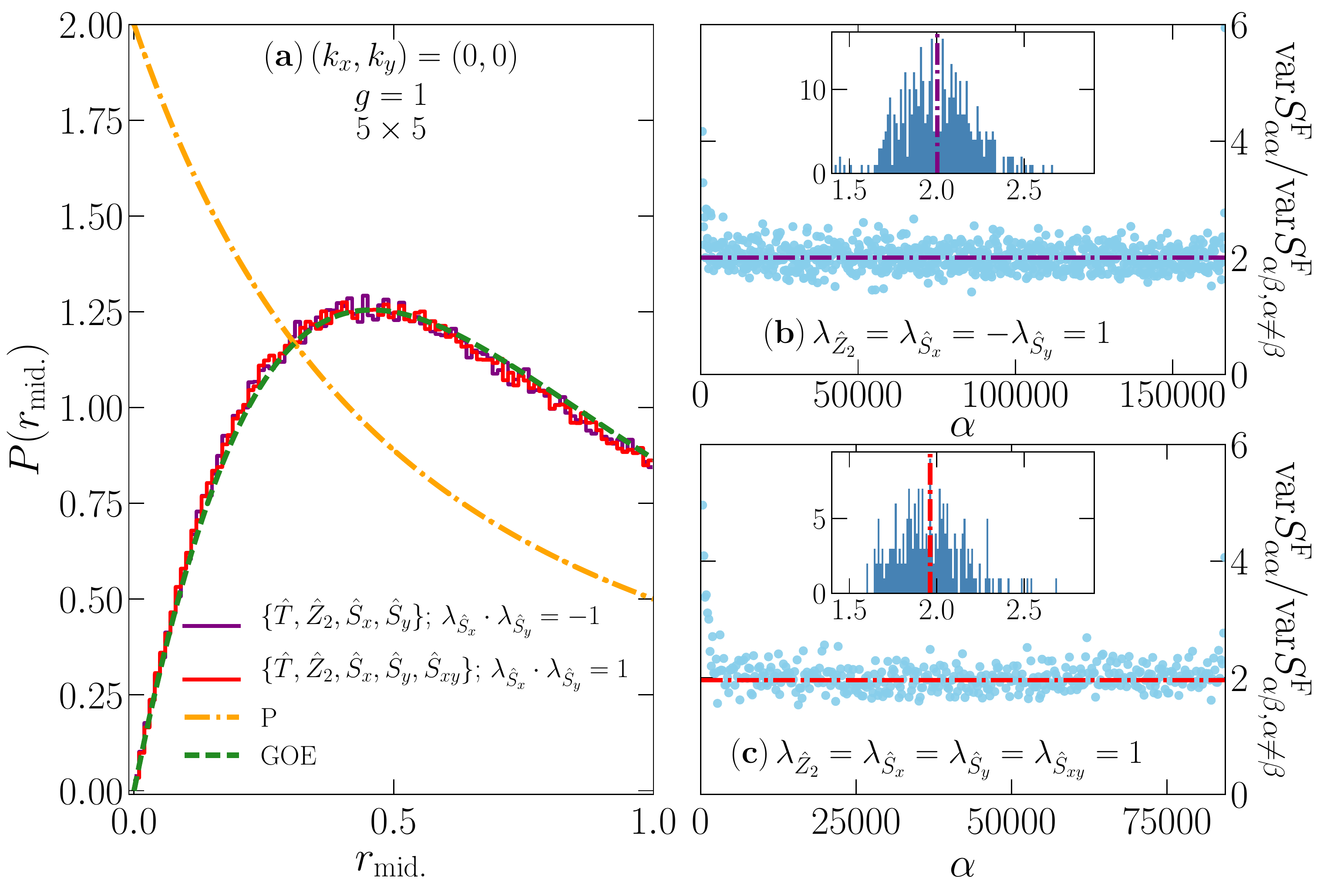}
 \vspace{-0.1cm}
 \caption{(Color online) Quantum chaos analysis and connection to the GOE predictions in a $5\times 5$ lattice for $g = 1$. (a) Probability distribution of the ratio of adjacent energy gaps in the central one half of the spectrum. The results are averaged over equivalent subsectors with $\hat S_x$ and $\hat S_y$ parities obeying $\lambda_{\hat S_x}\cdot\lambda_{\hat S_y} = -1$ or $\lambda_{\hat S_x}\cdot\lambda_{\hat S_y} = 1$ (see text and Appendix~\ref{sec:appendix_split}). The dashed (dashed-dotted) line depicts the GOE (Poisson distribution) prediction. The Poisson distribution prediction, describing uncorrelated eigenenergy levels, is $P_\text{P}(r)=2\Theta(1-r)/(1+r^2)$~\cite{atas_bogomolny_13}. (b, c) Ratio of variances of the diagonal and off-diagonal matrix elements of the structure factor as a function of the energy eigenstate number (ordered with increasing energy) for a subsector with $\lambda_{\hat S_x}\cdot\lambda_{\hat S_y} = -1$ (b) and $\lambda_{\hat S_x}\cdot\lambda_{\hat S_y} = 1$ (c). The horizontal dashed-dotted lines show the average of the ratio of variances considering eigenstates in the central one half of the spectrum. (Insets) Distribution of the ratios of variances in the central one half of the spectrum and the average (vertical dashed-dotted lines). The windows used to compute the ratios of variances contain 200 energy eigenstates.}
 \label{fig:P_r_mid_ratio_var_ns_25}
\end{figure}

\begin{figure}[!tb] 
  \includegraphics[width=0.95\columnwidth]{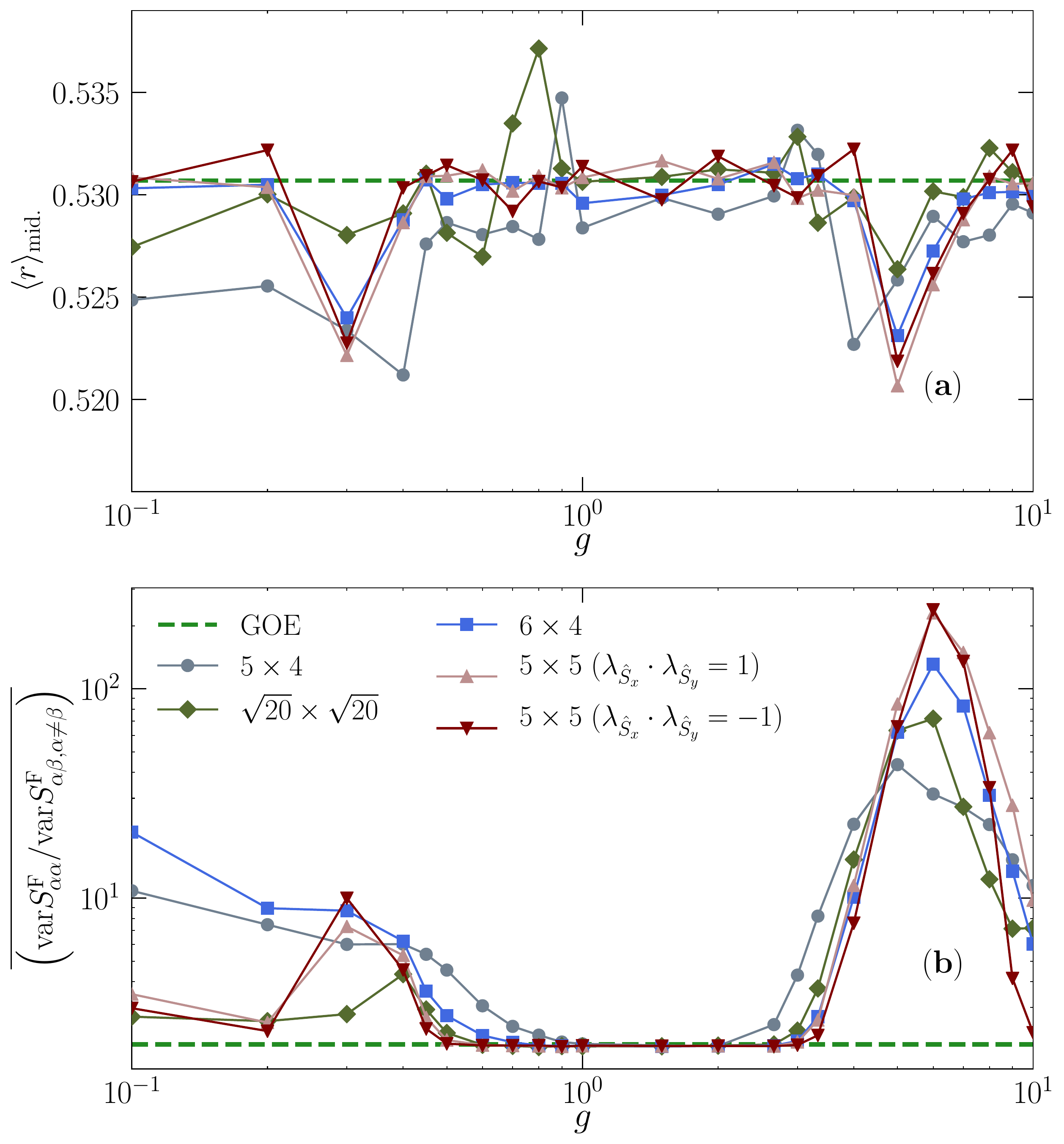}
 \vspace{-0.1cm}
 \caption{(Color online) (a) Average ratio of adjacent energy gaps in the central one half of the spectrum $\langle r\rangle_{ \text{mid.}}$, for four lattice sizes, as a function of the strength of the transverse field. The results for $\langle r\rangle_{ \text{mid.}}$ are obtained averaging over equivalent symmetry irreducible subsectors in each lattice. (b) Average ratio of variances of the diagonal and off-diagonal matrix elements of the structure factor, computed in the central one half of the spectrum, as a function of $g$. The ratios of variances are calculated using windows containing 50 energy eigenstates, and, like in (a), the results reported are the average over equivalent symmetry irreducible subsectors. The dashed lines depict the predictions of the GOE. Only the zero momentum subsector is considered for the regular lattices (as in Fig.~\ref{fig:P_r_mid_ratio_var_ns_25}), while $\vec k = \frac{\pi}{5}(2,1)$ is used for the 20-sites tilted lattice. The results reported for the latter are the average over the two $\hat Z_2$ subsectors.}
 \label{fig:r_mid_and_ratio_var_vs_g}
\end{figure}

Figure~\ref{fig:P_r_mid_ratio_var_ns_25}(a) shows the distribution of the ratio of adjacent gaps obtained in the central one half of the spectrum~\footnote{Since the 2D-TFIM [Eq.~(\ref{eq:hamiltonian})] only has short-ranged interactions and no randomness, the random matrix theory predictions for the level spacing statistics only apply away from the edges of the spectrum~\cite{brody_flores_81, kaplan2000wave, flores2001spectral, santos_rigol_10a, santos_rigol_10b}.}, averaged between the equivalent symmetry irreducible subsectors with either $\lambda_{\hat S_x}\cdot\lambda_{\hat S_y} = -1$ or $\lambda_{\hat S_x}\cdot\lambda_{\hat S_y} = 1$ in the $5\times5$ lattice. The former contain ${\cal D}^ \prime \approx 167000$ states, while the extra $\hat S_{xy}$ mirror symmetry in the latter results in subsectors with ${\cal D}^ \prime \approx 85000$ states (see Appendix A). For Hamiltonians that are time-reversal symmetric, the appropriate random matrix ensemble is the GOE, whose distribution of the ratio of adjacent gaps is: $P_\text{GOE}(r) = (27/4) [(r+ r^2)\Theta(1-r)] / (1+r+r^2)^{5/2}$~\cite{atas_bogomolny_13}. This prediction is depicted by the (green) dashed line in Fig.~\ref{fig:P_r_mid_ratio_var_ns_25}, and it is in almost perfect agreement with the distribution of the adjacent energy gaps obtained numerically. From this, one can conclude that quantum chaotic behavior takes place in the 2D-TFIM. The small differences seen between the analytic predictions and the numerical results are due to the fact that the former are exact only for $3\times3$ matrices~\cite{atas_bogomolny_13,mondaini_fratus_16}.

In Fig.~\ref{fig:r_mid_and_ratio_var_vs_g}(a), we show numerical results for the average of adjacent energy gaps in the central one half of the spectrum, $\langle r\rangle_{ \text{mid.}}$, as a function of the transverse field $g$ for four lattice sizes. The prediction from the GOE is $\langle r\rangle_\text{GOE} \approx 0.5359$~\cite{atas_bogomolny_13}, and is depicted in Fig.~\ref{fig:r_mid_and_ratio_var_vs_g}(a) as a horizontal dashed line. Even for the smallest lattice sizes studied, the average ratio of adjacent gaps is quite close to $\langle r\rangle_\text{GOE}$ for $g\simeq J$. With increasing lattice size one can see that the agreement improves for $g\simeq J$ and extends toward $g\ll1$ and $g\gg1$, which suggests that any nonzero but finite value of $g$ results in quantum chaotic behavior in the thermodynamic limit \cite{santos_rigol_10a, rigol_santos_10}.

\subsection{Ratio of diagonal and off-diagonal variances of matrix elements of observables}\label{sec:vratios}

Given the quantum chaotic behavior observed in the energy spectrum, one might wonder whether other properties of random matrices are present in the 2D-TFIM. Of particular relevance to eigenstate thermalization, one can show that the variance of the diagonal and off-diagonal matrix elements of Hermitian operators in the eigenstates of random matrices are proportional to each other. For the GOE, the proportionality constant is exactly 2~\cite{dalessio_kafri_16}.

This follows from the fact that the eigenstates of real symmetric random matrices are essentially orthonormal random vectors in arbitrary bases. Let us imagine we have a Hermitian operator $\hat A$, with $\hat A|i\rangle=A_i|i\rangle$. The matrix elements of $\hat A$ in the eigenkets of a real symmetric random matrix $\{|\alpha\rangle\}$ read
\begin{equation}
 A_{\alpha\beta} \equiv \langle \alpha|\hat A|\beta\rangle = \sum_{i,j}\langle\alpha|i\rangle\langle i|\hat A|j\rangle\langle j|\beta\rangle  = \sum_i A_i c_i^\alpha c_i^\beta,\nonumber
\end{equation}
where we defined $c_i^\alpha\equiv\langle i|\alpha\rangle = \langle \alpha| i\rangle$. The $c_i^\alpha$'s are Gaussian distributed with zero mean and variance equal to $1/{\cal D}$ (to leading order), where ${\cal D}$ is the dimension of the random matrix. Two results follow immediately from this: (1) $\overline{c_i^\alpha c_j^\beta} = (1/{\cal D})\delta_{\alpha\beta}\delta_{ij}$, and (2) different moments of the distribution of $c_i^\alpha$'s are related, e.g., $\overline{(c_i^\alpha)^4}=3\overline{(c_i^\alpha)^2}$. 

We are interested in the fluctuations of the diagonal and off-diagonal matrix elements of operator $\hat A$ in the eigenstates of a random matrix: $\text{var\,} A_{\alpha\alpha} = \overline{A_{\alpha\alpha}^2} - \overline{A_{\alpha\alpha}}^2$ and $\text{var\,} A_{\alpha\beta}^{(\alpha\neq\beta)} = \overline{A_{\alpha\beta}^2} - \overline{A_{\alpha\beta}}^2$, respectively. Using the two results mentioned above, one gets
\begin{eqnarray}
 \text{var\,}A_{\alpha\alpha} &=& \sum_{i,j}A_i A_j\overline{c_i^\alpha c_i^\alpha c_j^\alpha c_j^\alpha}-\sum_{i,j}A_i A_j\overline{c_i^\alpha c_i^\alpha}\,\overline{c_j^\alpha c_j^\alpha} \nonumber \\ &=&\sum_{i}A_i^2\left[\overline{(c_i^\alpha)^4}-\overline{(c_i^\alpha)^2}^2\right]=\frac{2}{{\cal D}^2}\sum_i A_i^2,
\end{eqnarray}
and 
\begin{eqnarray}
 \text{var\,}A_{\alpha\beta}^{(\alpha\neq\beta)} &=& \sum_{i,j}A_i A_j\overline{c_i^\alpha c_i^\beta c_j^\alpha c_j^\beta}-\sum_{i,j}A_i A_j\overline{c_i^\alpha c_i^\beta}\,\overline{c_j^\alpha c_j^\beta} \nonumber \\
 &=&\sum_{i}A_i^2\overline{(c_i^\alpha)^2(c_i^\beta)^2}-0=\frac{1}{{\cal D}^2}\sum_i A_i^2,
\end{eqnarray}
which means that the ratio between the variance of the diagonal and off-diagonal matrix elements of $\hat A$ is
\begin{equation}
 \frac{\text{var\,}A_{\alpha\alpha}}{\text{var\,}A_{\alpha\beta}^{(\alpha\neq\beta)}}=2.
 \label{eq:ratio_of_var}
\end{equation}

This is consistent with the ETH ansatz, as in the latter the eigenstate to eigenstate fluctuations of the diagonal matrix elements of observables are exponentially small in the system size, as the off-diagonal matrix elements are. The question that remains is whether the ratio of the variances of diagonal and off-diagonal matrix elements of observables in physical systems with short-range interactions and no randomness is constant away from the edges of the spectrum and equal to 2. Since in such systems the diagonal and off-diagonal matrix elements of observables are expected to have structure in $\bar E$ and $\omega$ [see the ETH ansatz \eqref{eq:eth_ansatz}] the calculation of the variances has to be carried out within sufficiently small energy windows so that ${\cal O}(\bar E)$ and $e^{-S(\bar E)/2}f_{O}(\bar E,\omega)$ are essentially constant (as the corresponding terms are in random matrices~\cite{dalessio_kafri_16}).

To address this question, we first study the matrix elements of the ferromagnetic structure factor,
\begin{equation}
 \hat S^\text{F} \equiv \frac{1}{N}\sum_{{\bf i},{\bf j}}{\hat \sigma_{\bf i}^z}{\hat \sigma_{\bf j}^z}.
\end{equation}
This non-local few-body observable is an order parameter for the phase transitions that occur in the 2D-TFIM. Its expectation value is extensive (order 1) in the ordered (paramagnetic) phase. Since the focus of our study are energy eigenstates in the central one half of the spectrum (``high temperature'' eigenstates; see Appendix~\ref{sec:mbdos}), the eigenstate expectation values of $\hat S^\text{F}$ in our calculations are $O(1)$~\cite{mondaini_fratus_16}. 

In Fig.~\ref{fig:P_r_mid_ratio_var_ns_25}(b) and \ref{fig:P_r_mid_ratio_var_ns_25}(c), we show the ratio of the variances of the diagonal and off-diagonal matrix elements of the structure factor in the energy eigenstates of the 2D-TFIM, as a function of the eigenstate index. The results were obtained on the $5\times5$ lattice, within two symmetry irreducible subsectors with $\lambda_{\hat Z_2} = \lambda_{\hat S_x}=-\lambda_{\hat S_y}=1$ [Fig.~\ref{fig:P_r_mid_ratio_var_ns_25}(b)] and $\lambda_{\hat Z_2} = \lambda_{\hat S_x}=\lambda_{\hat S_y}=\lambda_{\hat S_{xy}}=1$ [Fig.~\ref{fig:P_r_mid_ratio_var_ns_25}(c)], for $g=1$. It is remarkable that, away from the edges of the spectrum, the ratios of variances fluctuate about 2. Actually, the average of the ratios of variances within the central one half of the spectrum, shown as horizontal dashed-dotted lines, is very close to 2 (closer for the largest symmetry irreducible subsector). The insets in Figs.~\ref{fig:P_r_mid_ratio_var_ns_25}(b) and \ref{fig:P_r_mid_ratio_var_ns_25}(c) show the distribution of the ratios of variances when considering, once again, only the central one half of the spectrum (vertical dashed-dotted lines depict the average).

In Figs.~\ref{fig:r_mid_and_ratio_var_vs_g}(b), we plot the average ratio of variances of the diagonal and off-diagonal matrix elements of the structure factor as a function of the transverse field strength, for the same lattice sizes as in Fig.~\ref{fig:r_mid_and_ratio_var_vs_g}(a). The correlation between the values of $g$ for which the average ratio of variances is closest to 2 [Fig.~\ref{fig:r_mid_and_ratio_var_vs_g}(b)] and for which the average ratio of adjacent energy gaps is closest to the GOE prediction [Fig.~\ref{fig:r_mid_and_ratio_var_vs_g}(a)] is apparent. With increasing system size, one can see that the range of values of $g$ over which the numerical results are closest to random matrix theory prediction increases. We note that, when departing from the GOE predictions, the average ratio of variances increases. This is the result of an increase in the eigenstate-to-eigenstate fluctuations of the diagonal matrix elements~\cite{rigol_09a, rigol_09b}, and an increase of the number of off-diagonal matrix elements that become very small~\cite{rigol_09b, khatami_pupillo_13}. The latter effect can make the ratio of variances become arbitrarily large.

Our results contrast the ones obtained by Steinigeweg \textit{et al.}~\cite{steinigeweg_herbrych_13} for the ratios of variances of the diagonal and off-diagonal matrix elements of current operators in nonintegrable fermionic chains. They were found to exhibit a significant dependence on the energy of the eigenstates, to differ from 2 in the center of the energy spectrum, and to differ between the spin and energy currents. In our calculations, we have found that lack of agreement with the random matrix theory prediction can be a result of finite-size effects and/or the width of the windows used to compute the variances. Finite-size effects affect different observables in different ways, and the size of the windows used to compute the variances needs to be selected with care to avoid the influence of the nontrivial structure of matrix elements of observables in physical Hamiltonians (not present in random matrix theory).

\begin{figure}[!tb] 
  \includegraphics[width=0.95\columnwidth]{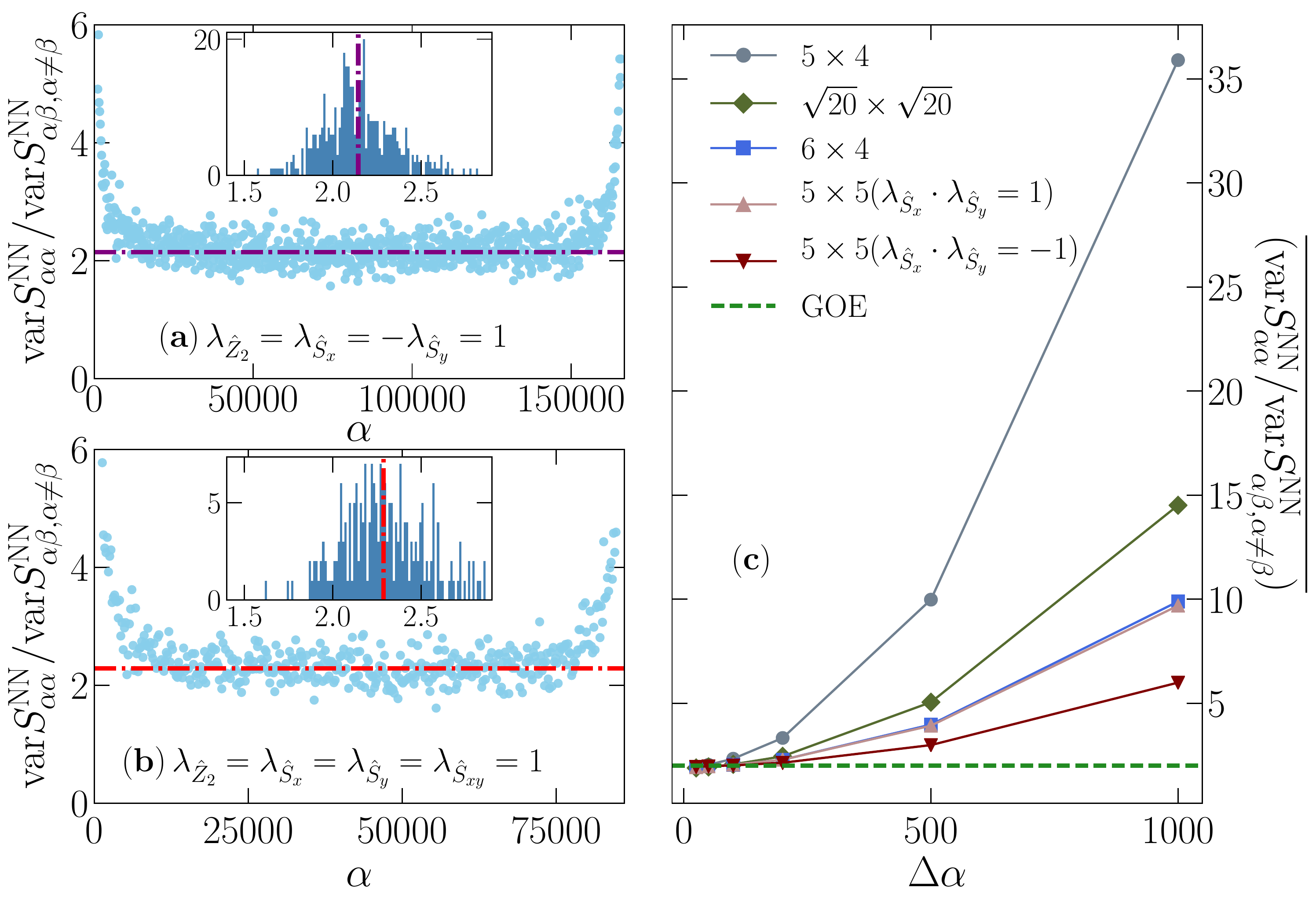}
 \vspace{-0.1cm}
 \caption{(Color online) (a, b) Same as Figs.~\ref{fig:P_r_mid_ratio_var_ns_25}(b) and~\ref{fig:P_r_mid_ratio_var_ns_25}(c) but for the nearest-neighbor spin-spin correlation function. (c) Average ratio of variances for the nearest-neighbor spin-spin correlation function, computed in the central one half of the spectrum, as a function of the number of eigenstates in the windows. Results are shown for one of the largest symmetry irreducible subsectors for four lattice sizes.}
 \label{fig:ratio_of_var_sisj_nn}
\end{figure}

In Figs.~\ref{fig:ratio_of_var_sisj_nn}(a) and~\ref{fig:ratio_of_var_sisj_nn}(b), we show results of calculations identical to those reported in Figs.~\ref{fig:P_r_mid_ratio_var_ns_25}(b) and~\ref{fig:P_r_mid_ratio_var_ns_25}(c) but for the nearest-neighbor spin-spin correlation function, 
\begin{equation}
 \hat S^\text{NN} \equiv \frac{1}{N}\sum_{\langle {\bf i},{\bf j}\rangle}{\hat \sigma_{\bf i}^z}{\hat \sigma_{\bf j}^z}. 
\end{equation}
$\hat S^\text{NN}$ exhibits larger finite-size effects than the structure factor, as apparent from the fact that for $\hat S^\text{NN}$: (1) the fluctuations of the ratio of variances are slightly larger throughout the spectrum, and (2) the average exhibits a larger departure from 2. In Fig.~\ref{fig:ratio_of_var_sisj_nn}(c), we show that if one increases the size of the windows used to compute the variances, in order to reduce the fluctuations of their ratio, then the nontrivial structure of the diagonal and off-diagonal matrix elements of observables kicks in and the average ratio of variances departs from the GOE prediction. The departure depends strongly on the lattice geometries used. See Appendix~\ref{sec:ratio_of_var} for the equivalent of  Fig.~\ref{fig:ratio_of_var_sisj_nn}(c) for the ferromagnetic structure factor.

In Appendix~\ref{sec:diag_mat_elem}, we show results for the eigenstate expectation values of $\hat S^\text{F}$ and $\hat S^\text{NN}$ for different values of $g$ and system sizes. They further help gaining an understanding of finite-size effects in different observables, as well as how ETH sets in when $g$ is increased from zero and breaks down as $g$ becomes much larger than $J$.

\section{Off-diagonal matrix elements and the ETH ansatz}\label{sec:sec4}

\begin{figure}[!b] 
  \includegraphics[width=0.95\columnwidth]{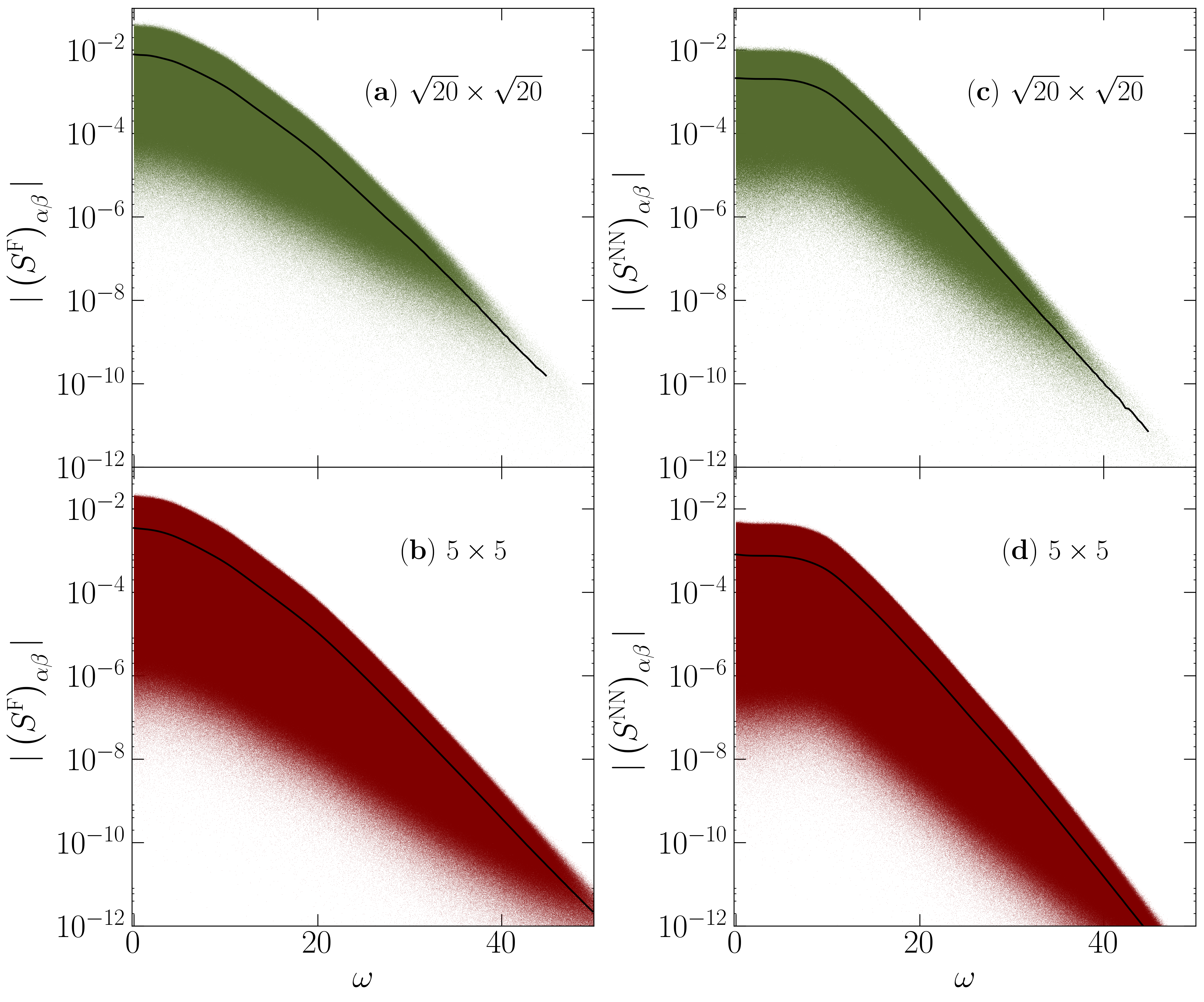}
 \vspace{-0.1cm}
 \caption{(Color online) Absolute value of the off-diagonal matrix elements of the structure factor (a, b) and of the nearest neighbor spin-spin correlation function (c, d), for $2|\bar E|/N\leq0.1$, plotted as a function of $\omega$ for $g=1$. Panels (a) and (c) show results for the $\sqrt{20}\times\sqrt{20}$ lattice, and panels (b) and (d) show results for the $5\times 5$ lattice. The matrix elements were obtained in the subsector with $\lambda_{\hat Z_2} = 1$ for the 20-sites tilted lattice, and in the largest symmetry irreducible subsector with $\lambda_{\hat Z_2} = \lambda_{\hat S_x} = -\lambda_{\hat S_y}=1$ for the 25-sites lattice. The continuous lines are running averages. Because of the large number of matrix elements present in the $5\times 5$ lattice, in panels (b) and (d) we plot only every second one.}
 \label{fig:s_alpha_beta_large_5_times_5}
\end{figure}

Having unveiled random matrix theory behavior in the matrix elements of observables in microscopic energy windows, we now study the off-diagonal matrix elements of the structure factor and the nearest-neighbor spin-spin correlation function as a function of $\omega = E_\alpha - E_\beta$. 

Figure~\ref{fig:s_alpha_beta_large_5_times_5} shows the absolute value of the off-diagonal matrix elements of the structure factor [Figs.~\ref{fig:s_alpha_beta_large_5_times_5}(a) and~\ref{fig:s_alpha_beta_large_5_times_5}(b)] and of the nearest-neighbor spin-spin correlation function [Figs.~\ref{fig:s_alpha_beta_large_5_times_5}(c) and~\ref{fig:s_alpha_beta_large_5_times_5}(d)] versus $\omega$, for $\omega>0$ ($S^{\rm {F}}_{\alpha\beta}$ and $S^{\rm NN}_{\alpha\beta}$ are symmetric) and $2|\bar E|/N=|E_\alpha+E_\beta|/N\leq0.1$. We report results for the two largest lattices with square aspect ratio, namely, the $\sqrt{20}\times\sqrt{20}$ lattice [Figs.~\ref{fig:s_alpha_beta_large_5_times_5}(a) and~\ref{fig:s_alpha_beta_large_5_times_5}(c)] and the $5\times5$ lattice [Figs.~\ref{fig:s_alpha_beta_large_5_times_5}(b) and~\ref{fig:s_alpha_beta_large_5_times_5}(d)]. The off-diagonal matrix elements of each observable are qualitatively similar in the two lattices.

In order to study the behavior of the smooth $e^{-S(\bar E)/2}f_{O}(\bar E,\omega)$ function [see Eq.~\eqref{eq:eth_ansatz}], we compute the running (or coarse-grained) average over small windows of width $\delta\omega$ for the two observables of interest. The widths of the windows are different for different lattices. They are selected such that the result of the averaging produces a smooth curve that is not sensitive to the exact value of $\delta\omega$ chosen. The results of such a coarse-graining procedure are reported in Fig.~\ref{fig:s_alpha_beta_large_5_times_5} as continuous black lines.

\begin{figure}[!tb] 
  \includegraphics[width=0.95\columnwidth]{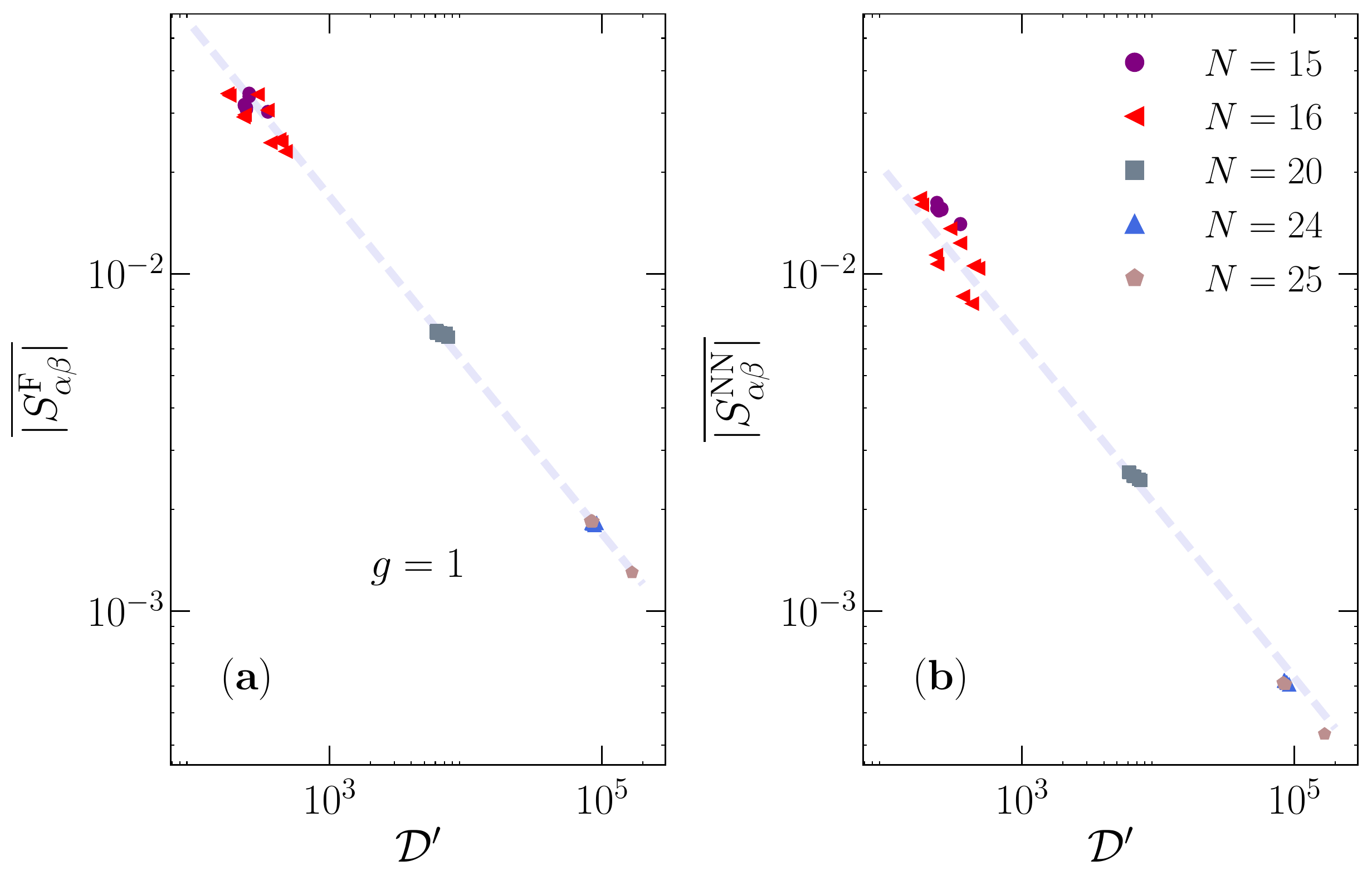}
 \vspace{-0.1cm}
 \caption{(Color online) Average of the absolute value of the off-diagonal matrix elements of the structure factor (a) and the nearest neighbor spin-spin correlation function (b) as a function of the size of the symmetry irreducible subsectors on various regular lattices for $g=1$. The matrix elements included in the average are for eigenstates satisfying $2|\bar E|/N\leq0.1$. The dashed lines depict a fit of the results in lattices with $N=20$, 24 and 25 to a constant times $({\cal D}^\prime)^{-1/2}$.}
 \label{fig:gamma_vs_D}
\end{figure}

Since we have chosen $\bar E/N\approx0$, the energy at infinite temperature, $e^{-S(\bar E)/2}$ is nothing but $({\cal D}^\prime)^{-1/2}$ where ${\cal D}^\prime$ is the size of the subsector studied. That the off-diagonal matrix elements in our calculations are indeed proportional to $({\cal D}^\prime)^{-1/2}$ (and, hence, exponentially small in the system size) can be verified by computing the average
\begin{equation}
\label{eq:ave}
\overline{|O_{\alpha\beta}|} = \frac{1}{{\cal N}}\sum_{\substack{\alpha,\beta \\ (\alpha\neq\beta)}} |O_{\alpha\beta}|,
\end{equation}
where ${\cal N}$ is the number of terms contributing to the sum, and plotting it vs ${\cal D}^\prime$ for different lattice sizes and, within a given lattice size, for different subsectors. In Fig.~\ref{fig:gamma_vs_D}, we present such a plot for the average between the off-diagonal matrix elements of eigenstates with $2|\bar E|/N\leq0.1$, in all subsectors of the zero momentum sector of the regular lattices studied, and for $g=1$. The numerical results exhibit an excellent agreement with the expected $({\cal D}^\prime)^{-1/2}$ behavior. This allows us to extract $f_{O}(\bar E,\omega)$, up to a constant, from the running averages.

\begin{figure}[!tb] 
  \includegraphics[width=0.99\columnwidth]{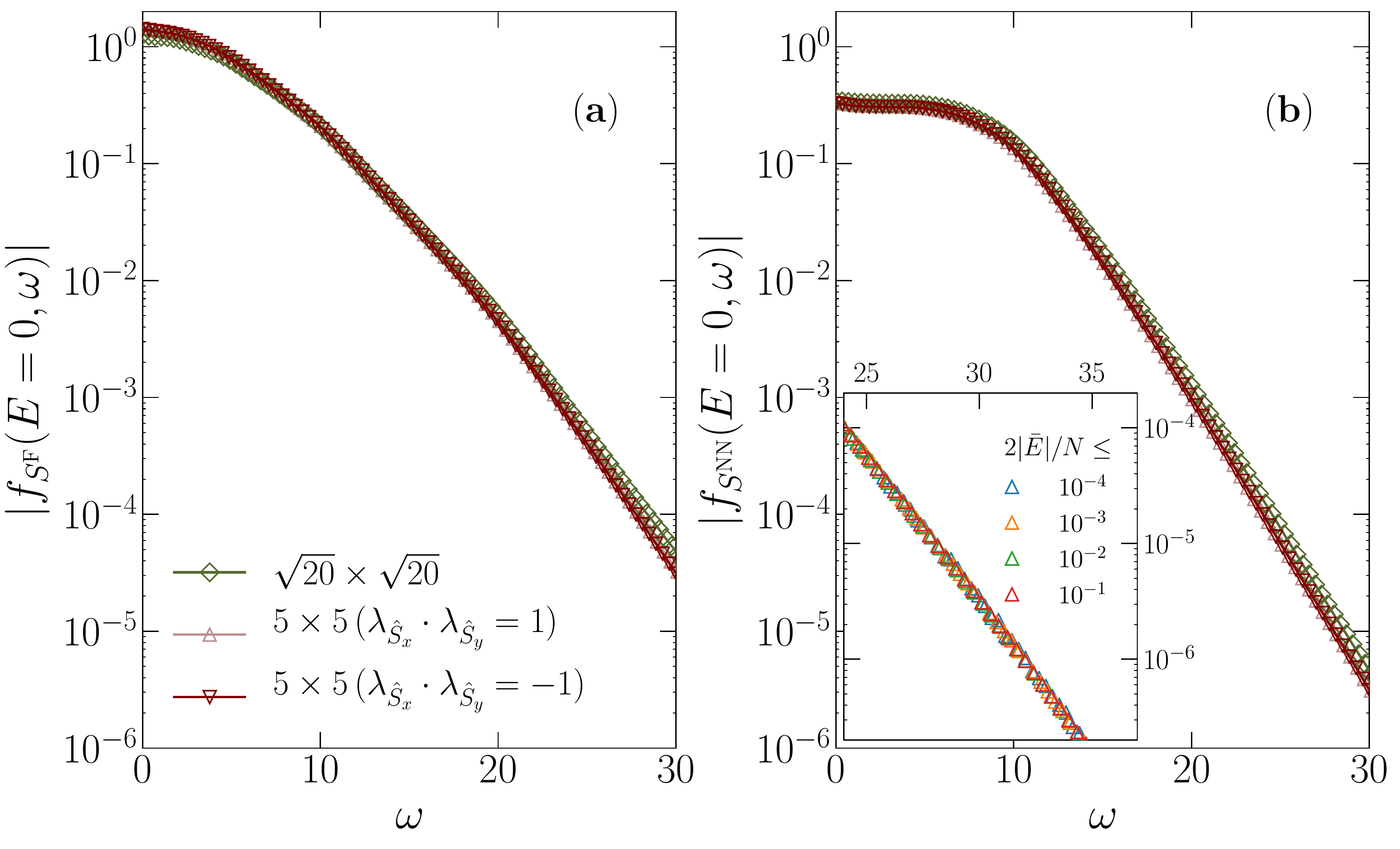}
 \vspace{-0.4cm}
 \caption{(Color online) Smooth function $f_{O}(\bar E,\omega)$ in the ETH ansatz \eqref{eq:eth_ansatz}, (a) $f_{S^{\rm {F}}}(\bar E\approx0,\omega)$ and (b) $f_{S^{\rm {NN}}}(\bar E\approx0,\omega)$, plotted versus $\omega$. $f_{S^{\rm {F}}}(\bar E\approx0,\omega)$ and $f_{S^{\rm {NN}}}(\bar E\approx0,\omega)$ are obtained from the results reported in Fig.~\ref{fig:s_alpha_beta_large_5_times_5} as explained in the text. The inset in panel (b) depicts $f_{S^{\rm {NN}}}(\bar E\approx0,\omega)$ for $2|\bar E|/N\leq10^{-4},\,10^{-3},\,10^{-2}$ and $10^{-1}$ for the $5\times 5$ lattice in the subsector with $\lambda_{\hat Z_2} = \lambda_{\hat S_x} = \lambda_{\hat S_y}=\lambda_{\hat S_{xy}}=1$.}
 \label{fig:f_function_different_sectors_asp_rat_square}
\end{figure}

Figure~\ref{fig:f_function_different_sectors_asp_rat_square} shows $f_{O}(\bar E,\omega)$ determined this way, for the structure factor [Fig.~\ref{fig:f_function_different_sectors_asp_rat_square}(a)] and for the nearest-neighbor spin-spin correlation function [Fig.~\ref{fig:f_function_different_sectors_asp_rat_square}(b)], for one subsector of the 20-sites tilted lattice and two subsectors of the $5\times5$ lattice. The results for each observable in the two lattices are very close to each other. For both observables, $f_{O}(\bar E\approx0,\omega)$ is nearly constant for very small values of $\omega$ (this is the regime in which the random matrix predictions were tested in Sec.~\ref{sec:vratios}) and decays rapidly for large $\omega$. In the latter regime, the decay of $f_{S^{\rm {F}}}(\bar E\approx0,\omega)$ appears to be a composition of several exponentials, while the one of $f_{S^{\rm {NN}}}(\bar E\approx0,\omega)$ is close to that of a single exponential. Exponential decays of $f_{O}(\bar E,\omega)$ for local observables $\hat O$ at large $\omega$ have been observed in previous works~\cite{khatami_pupillo_13, beugeling_moessner_15, dalessio_kafri_16}. They can be understood in terms of perturbation theory for systems with a bounded spectrum. The reason is that many-particle processes, which are suppressed exponentially, are required to connect eigenstates with large energy differences~\cite{dalessio_kafri_16}. The inset in Fig.~\ref{fig:f_function_different_sectors_asp_rat_square}(b) shows that the exponential behavior found in the $5\times 5$ lattice is robust to changes in the size of the window used in the calculations.

\section{Summary}\label{sec:sec5}
Using large (within full exact-diagonalization calculations) Hilbert space sizes, we studied properties of the matrix elements of two few-body observables in the eigenstates of the 2D-TFIM. We showed that the onset of quantum chaos, identified using a level spacing analysis, results in the applicability of another random matrix theory prediction. The ratio of the variances of the diagonal and off-diagonal matrix elements of observables, calculated within very small energy windows, is constant across the spectrum (excluding the edges) and equal to 2. We also studied the behavior of the off-diagonal matrix elements as a function of $\omega = E_\alpha - E_\beta$. We showed that, in the quantum chaotic regime, their smooth part can be well described by a function $e^{-S(\bar E)/2}f_{O}(\bar E, \omega)$ as prescribed by the ETH ansatz. $f_{O}(\bar E, \omega)$ was shown to be nearly constant for small values of $\omega$ and to exhibit a rapid (exponential) decay for large values of $\omega$. An interesting problem left for future studies is correlating the temporal evolution of observables at short, intermediate, and long times with the large, intermediate, and small $\omega$ behavior of $f_{O}(\bar E, \omega)$, and unveiling the effect of the initial states in the dynamics at different time scales.

\begin{acknowledgments}
RM is financially supported by the National Natural Science Foundation of China (NSFC) (Grant Nos. U1530401, 11674021 and 11650110441) and MR by the U.S. Office of Naval Research, Grant No. N00014-14-1-0540. The computations were performed in the Tianhe-2JK at the Beijing Computational Science Research Center (CSRC), the Institute for CyberScience at Penn State, and the Center for High-Performance Computing at the University of Southern California.
\end{acknowledgments}

\appendix

\section{Hilbert space subsectors}\label{sec:appendix_split}

We rewrite the original Fock basis, written in terms of the eigenstates of $\hat\sigma^z_{\bf i}$, using the following:
\\
\textit{\indent Translations} $\hat T$: $\hat T_{\vec R} |\psi_{\hat T_{\vec R}}\rangle = e^{-{\rm i}\vec k\cdot \vec R}|\psi_{\hat T_{\vec R}}\rangle$, where $\vec k$ defines the $N$ possible momentum sectors. For the regular lattices, we diagonalize only the zero momentum sector, i.e., we deal with real matrices. For the 20-sites tilted lattice, we diagonalize the $\vec k = \frac{\pi}{5}(2,1)$ momentum sector (see Ref.~\cite{mondaini_fratus_16}).
\\
\textit{\indent Spin-flip} $\hat Z_2$: $\hat Z_2 |\psi_{\hat T_{\vec R}, \hat Z_2}\rangle = \lambda_{\hat Z_2} |\psi_{\hat T_{\vec R},\hat Z_2}\rangle$, where $\lambda_{\hat Z_2} = \pm1$.
The remaining symmetries apply only to the zero momentum sector of the regular lattices.
\\
\textit{\indent Mirror in $x$, $\hat S_x$:} $\hat S_x |\psi_{\hat T_{\vec R}, \hat Z_2, \hat S_x}\rangle = \lambda_{\hat S_x} |\psi_{\hat T_{\vec R}, \hat Z_2, \hat S_x}\rangle$, where $ \lambda_{\hat S_x} = \pm1$.
\\
\textit{\indent Mirror in $y$, $\hat S_y$:} $\hat S_y |\psi_{\hat T_{\vec R}, \hat Z_2, \hat S_x, \hat S_y}\rangle = \lambda_{\hat S_y} |\psi_{\hat T_{\vec R}, \hat Z_2, \hat S_x, \hat S_y}\rangle$, where $\lambda_{\hat S_y} = \pm1$.

If $\ell_x=\ell_y$, then for $ \lambda_{\hat S_x} = \lambda_{\hat S_y}$ one also has a mirror symmetry along the $x=y$ line, $\hat S_{xy}$: $\hat S_{xy} |\psi_{\hat T_{\vec R}, \hat Z_2, \hat S_x, \hat S_y, \hat S_{xy}}\rangle = \lambda_{\hat S_{xy}} |\psi_{\hat T_{\vec R}, \hat Z_2, \hat S_x, \hat S_y, \hat S_{xy}}\rangle$, where $\lambda_{\hat S_{xy}} = \pm1$.

The splitting of sectors after each symmetry is given below for the two largest regular lattices that we study, namely, the $6\times4$ and $5\times5$ lattices. We start with the largest sector from translational symmetry, $\vec{k}=(0,0)$, and then apply $\hat Z_2$, $\hat S_x$, $\hat S_y$, and, if applicable $\hat S_{xy}$. The top (bottom) number in each curly brace refers to the size of the subsector with positive (negative) parity. 

For the $6\times4$ lattice the splitting of the sectors reads (ordered to show the splitting under $\hat Z_2$, $\hat S_x$, and $\hat S_y$):
\begin{equation}
\label{eq:splitting_f_sectors_6_x_4}
{\cal D}_{\vec k_{0}} = 699,600
\begin{cases} 
    350,064 \begin{cases} 181,012 \begin{cases} 93,202
                                             \\ 87,810 \end{cases}  \\
                      \\ 169,052 \begin{cases} 84,662
                                             \\84,390  \end{cases}

            \end{cases}\\
    349,536 \begin{cases} 180,232 \begin{cases} 91,652
                                             \\ 88,580 \end{cases} \\
                                             
                       \\  169,304   \begin{cases} 85,164
                                                  \\84,140 \end{cases}

            \end{cases}\\
\end{cases}\nonumber
\end{equation}
For the $5\times5$ lattice the splitting of the sectors reads (ordered to show the splitting under $\hat Z_2$, $\hat S_x$, $\hat S_y$, and, if applicable $\hat S_{xy}$):
\begin{equation*}
\label{eq:splitting_f_sectors_5_x_5}
{\cal D}_{\vec k_{0}} =1,342,208
\begin{cases} 
    671,104 \begin{cases} 337,192 \begin{cases} 170,440 \begin{cases} 86,056 \\ 
                                                                      84,384  \end{cases}
                                             \\ 166,752 \end{cases}  \\
                      \\ 333,912 \begin{cases} 166,752
                                             \\167,160 \begin{cases} 84,384 \\ 82,776 \end{cases}\end{cases}

            \end{cases}\\
    671,104 \begin{cases} 337,192 \begin{cases} 170,440 \begin{cases} 86,056 \\
                                                                      84,384\end{cases}
                                             \\ 166,752 \end{cases} \\
                                             
                       \\  333,912   \begin{cases} 166,752
                                             \\167,160 \begin{cases}  84,384 \\ 82,776 \end{cases}\end{cases}

            \end{cases}\\
\end{cases}\nonumber
\end{equation*}

\section{Many-body density of states}
\label{sec:mbdos}
\begin{figure}[!t] 
 \vspace{0.4cm}
  \includegraphics[width=0.97\columnwidth]{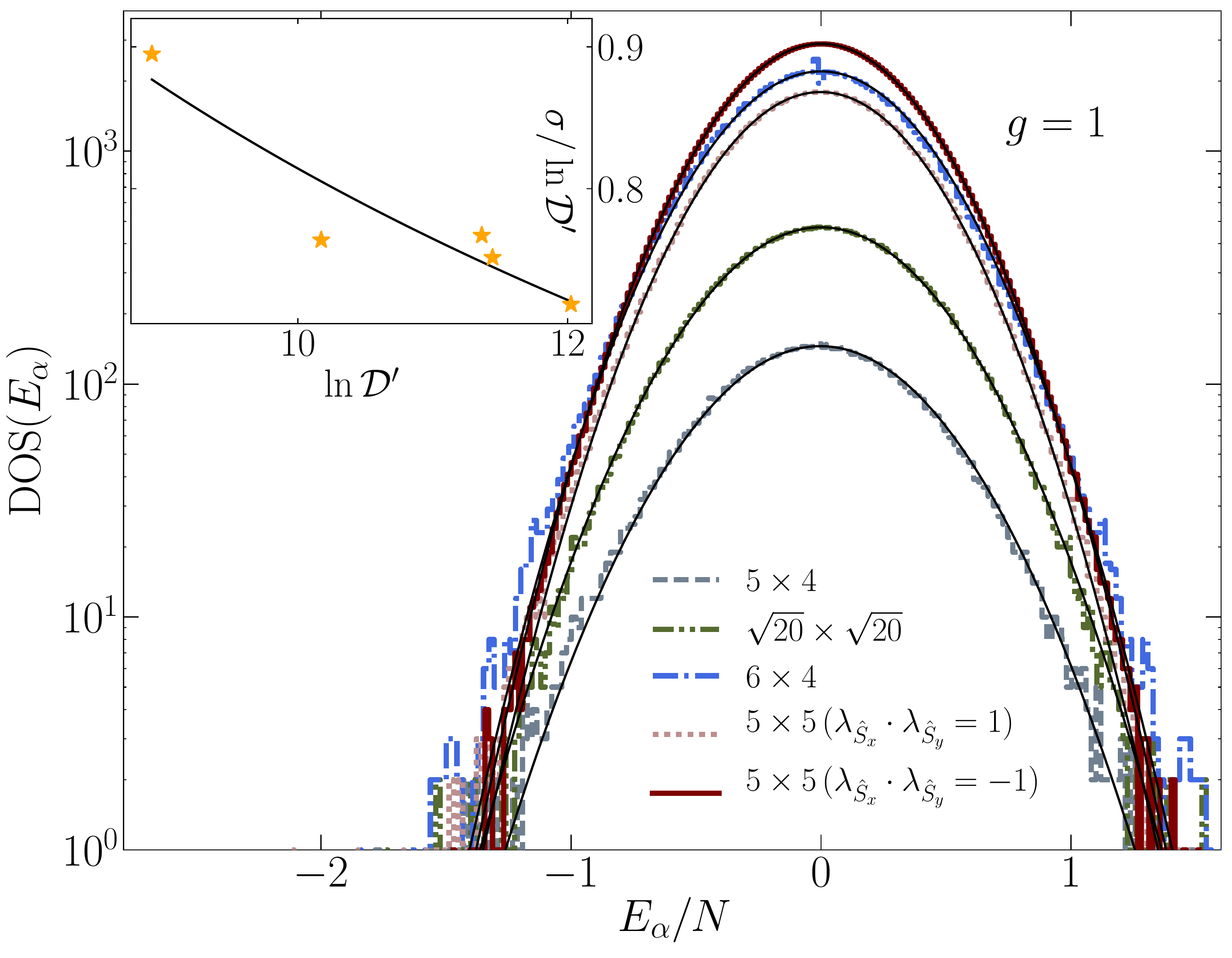}
  \vspace{-0.1cm}
 \caption{(Color online) Many-body density of states in the 2D-TFIM for subsectors that are even with respect to all parity operations (when applicable) for the lattices studied in this work, along with Gaussian fits. The inset depicts the ratio between the width of the Gaussian fits and $\ln {\cal D}^\prime$ vs $\ln {\cal D}^\prime$. A fit to a constant times $(\ln {\cal D}^\prime)^{-1/2}$ reveals the vanishing of that ratio  with increasing $\ln {\cal D}^\prime$ (which is proportional to $N$).}
 \label{fig:mbdosg1}
\end{figure}

The many-body density of states of systems with few-body interactions is, in general, Gaussian~\cite{brody_flores_81}. This is the case in the 2D-TFIM. The many-body density of states for subsectors that are even with respect to all parity operations (when applicable) for the lattices studied in this work are presented in Fig.~\ref{fig:mbdosg1} for $g=1$, along with Gaussian fits. The inset shows that the ratio between the width of the Gaussian fits and $\ln {\cal D}^\prime$ ($\ln {\cal D}^\prime$ is proportional to $N$) vanishes as $(\ln {\cal D}^\prime)^{-1/2}$. This means that, with increasing system size, the overwhelming majority of the eigenstates of the Hamiltonian have $E_\alpha/N$ increasingly close to 0, i.e., they are ``infinite temperature'' eigenstates. Those are the 2D-TFIM eigenstates studied in this work, as we have focused on eigenstates that are located in the central one half of the spectrum.

\section{Diagonal matrix elements}\label{sec:diag_mat_elem}

\begin{figure}[!b] 
  \includegraphics[width=0.97\columnwidth]{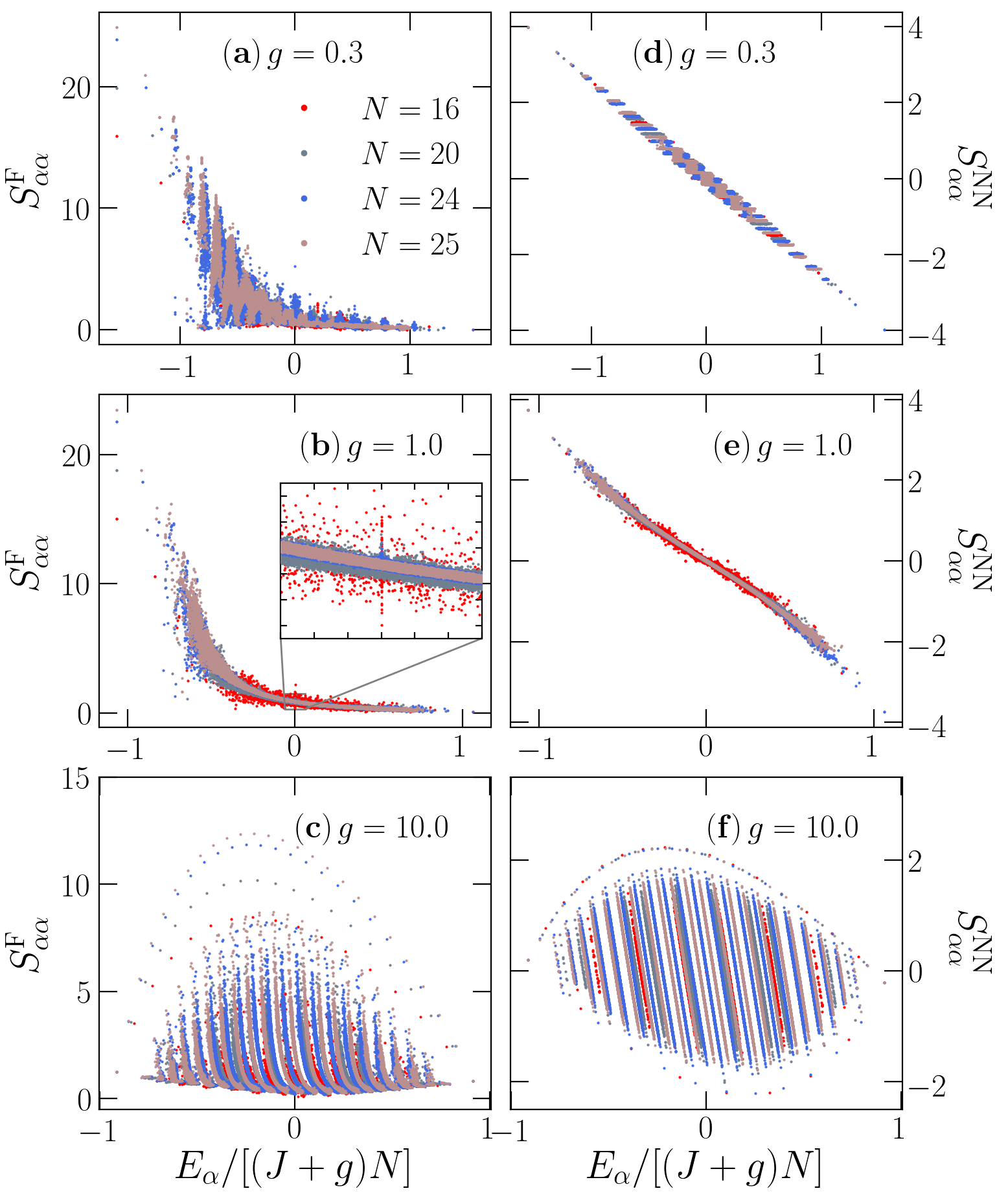}
 \vspace{-0.1cm}
 \caption{(Color online) Energy eigenstate expectation values of the ferromagnetic structure factor (nearest-neighbor spin-spin correlation function), $S^\text{F}_{\alpha\alpha}\equiv \langle\alpha|\hat S_\text{F}|\alpha\rangle$ $\left(S^\text{NN}_{\alpha\alpha}\equiv \langle\alpha|\hat S^\text{NN}|\alpha\rangle\right)$, plotted as a function of the eigenstate energies. Results are shown for all parity subsectors of the zero momentum sector in different regular lattices, and for three values of $g$. The inset in panel (b) highlights the occurrence of eigenstate thermalization through the narrowing of the support of the eigenstate expectation values with increasing system size.}
 \label{fig:O_alpha_alpha}
\end{figure}

In Fig.~\ref{fig:O_alpha_alpha}, we plot the expectation values of the structure factor [Fig.~\ref{fig:O_alpha_alpha}(a)--\ref{fig:O_alpha_alpha}(c)] and of the nearest-neighbor spin-spin correlation function [Fig.~\ref{fig:O_alpha_alpha}(d)--\ref{fig:O_alpha_alpha}(f)] in the eigenstates of the 2D-TFIM in four regular lattices. We show results from all parity subsectors of the zero momentum sector, for three values of $g$. As one can see in Figs.~\ref{fig:O_alpha_alpha}(a),~\ref{fig:O_alpha_alpha}(b),~\ref{fig:O_alpha_alpha}(d), and ~\ref{fig:O_alpha_alpha}(e), as $g$ departs from $g=0$ the expectation values of the observables become smooth functions of the energy (away from the edges of the spectrum), i.e., eigenstate thermalization occurs. If one further increases $g$, when $g\gg J$, the system approaches an integrable regime and eigenstate thermalization breaks down  [Figs.~\ref{fig:O_alpha_alpha}(c) and~\ref{fig:O_alpha_alpha}(f)].

\section{Ratio of variances for different energy windows}\label{sec:ratio_of_var}

\begin{figure}[!b] 
  \includegraphics[width=0.95\columnwidth]{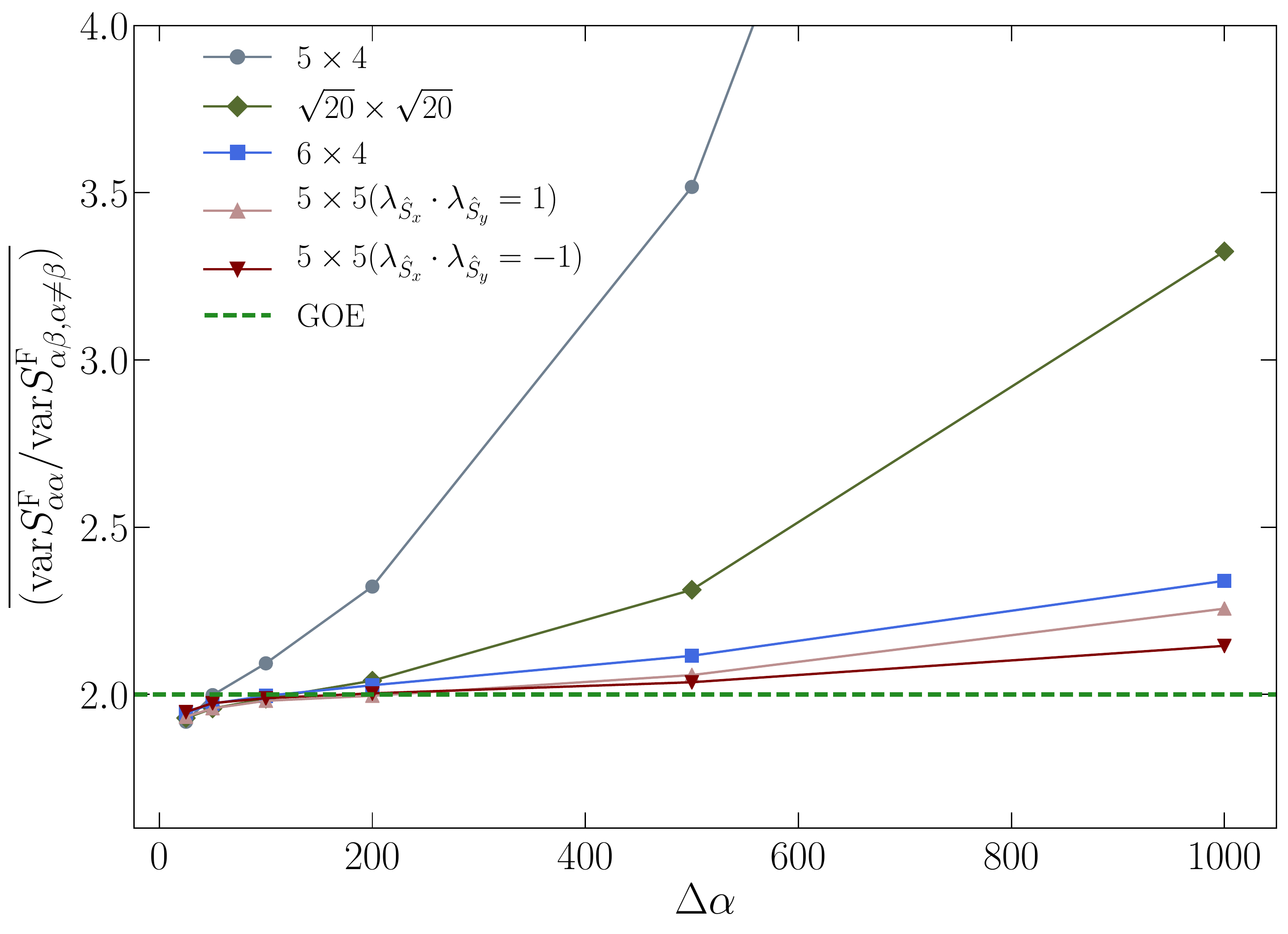}
 \vspace{-0.1cm}
 \caption{(Color online) Same as Fig.~\ref{fig:ratio_of_var_sisj_nn}(c) but for the ferromagnetic structure factor $S^\text{F}$.}
 \label{fig:ratio_of_var_sf}
\end{figure}

In Fig.~\ref{fig:ratio_of_var_sisj_nn}(c), we show how the ratio of variances of diagonal and off-diagonal matrix elements of the nearest-neighbor spin-spin correlation function depends on the size of the window used in the calculation. Figure~\ref{fig:ratio_of_var_sf} shows equivalent results for the ferromagnetic structure factor. The deviations of the results for the latter observable from the random matrix prediction exhibit a behavior that is qualitatively similar to that seen in Fig.~\ref{fig:ratio_of_var_sisj_nn}(c). However, for identical window sizes, the deviations of the results for the ferromagnetic structure factor are much smaller than the ones for the nearest-neighbor spin-spin correlation function. This is expected because, as mentioned before, the latter observable exhibits stronger finite-size effects in our calculations.

\bibliography{transverse_ising}

\end{document}